\documentclass[a4paper,10pt,twoside]{cpc-hepnp}

\usepackage{multicol}
\usepackage{graphicx}
\usepackage{booktabs}
\usepackage{amssymb,bm,mathrsfs,bbm,amscd}
\usepackage[tbtags]{amsmath}
\usepackage{lastpage}

\begin{document}

\fancyhead[co]{\footnotesize F. Jugeau: Hadronic linear potentials in $AdS$/CFT}


\title{Hadronic linear potentials in $AdS$/CFT\thanks{Invited talk at the 7$^{th}$ International Conference on Mathematical Methods in Physics, Joint Conference: CBPF-IMPA-ICTP-SISSA-TWAS, Rio de Janeiro, Brazil, 16-20 April 2012.}}

\author{%
Fr\'ed\'eric Jugeau%
} \maketitle

\address{%
Instituto de F\'isica, Universidade Federal do Rio de Janeiro, Cidade Universit\'aria, Ilha do Fund\~ao, Brazil\\
frederic.jugeau@if.ufrj.br
}

\begin{abstract}
In 1998, J. M. Maldacena conjectured a precise duality in the low-energy (or decoupling) regime between a large $N$ strongly-coupled $SU(N)$ super-Yang-Mills theory defined in the four-dimensional Minkowski world-volume $M^4$ of a stack of $N$ coincident D3-branes and the supergravity limit of a weakly-coupled type IIB closed superstring theory on $AdS_5\times S^5$. This communication aims at introducing concepts and methods used to derive, in the gauge/string correspondence framework, the interaction potentials of mesons and baryons at zero and finite temperature\cite{frederic}. Especially, emphasize will be made on the linear behavior of the bound-state potentials derived in the gravity side. Although $AdS$/CFT cannot be applied \emph{ab initio} to QCD-like gauge theories and remains a controversial topic, at least at zero temperature, we will also discuss $AdS$/QCD criteria proposed for a holographic description of the mechanism of confinement. 
\end{abstract}



\begin{multicols}{2}

\section{The area law in QCD and the heavy quark potentials in $AdS$/CFT}
A crucial breakthrough in the attempt to deal with strongly-coupled Yang-Mills theories came with the nowadays celebrated $AdS$/CFT correspondence whose the bulk-to-boundary dictionary has been established essentially in\cite{dictionary}. Shortly afterwards, the issue of calculating expectation values of Wilson loops was considered\cite{wilsonloop}, which is of significant importance since the Wilson loop, through the area law, consists of one of the most efficient tools for probing the large distance properties of confining QCD-like gauge theories\cite{arealaw}:
\begin{equation}\label{wilsonloop}
W[\mathcal{C}]\equiv\frac{1}{N}\langle Tr P e^{-\imath g\oint_{\mathcal{C}}A_{\mu}(x)dx^{\mu}}\rangle_A\underset{\mathcal{C}\rightarrow\infty}{=} e^{-\imath\sigma_t A[\mathcal{C}]}
\end{equation}   
with $A[\mathcal{C}]$ the area of the minimal surface of boundary $\mathcal{C}$. Then, the Feynman-Kac formula gives a linear static potential for an infinitely massive quark-antiquark pair:
\begin{equation}
V_{Q\bar{Q}}(r)=-\lim_{T\rightarrow\infty}\frac{1}{\imath T}\ln\,W[\mathcal{C}]=\sigma_t r\;.
\end{equation}
On the contrary, because of the underlying conformal symmetry on which $AdS$/CFT relies, the supergravity side computation of the infinitely heavy $Q\bar{Q}$ pair (non-dynamical external probes) potential shows instead a $1/r$ behavior for all distances $r$ and goes in terms of the 't Hooft coupling $\lambda$ as $\sqrt{\lambda}$, which reflects an intrinsically non-perturbative result\cite{wilsonloop}:
\begin{equation}\label{result}
V_{Q\bar{Q}}(r)=-\frac{4\sqrt{2}\pi^2}{\Gamma(1/4)^4}\frac{\sqrt{\lambda}}{r}\;.
\end{equation}
According to Maldacena's recipe, the expectation value of the \emph{spacetime} Wilson loop (\emph{spatial} loops - which follow the same prescription - will also be considered when deriving area law for $3d$ and $4d$ conformal field theories at finite temperature) is dual to the full partition function of the string theory which, in the low-energy supergrativy limit where the stringy effects are small, reduces to the proper area of the string world-sheet with the loop as the boundary. One might sketch the $AdS$/CFT steps of the calculation as follows:
\begin{equation}
W[\mathcal{C}]\underset{AdS\textrm{/CFT}}{\sim} Z_{string}[\mathcal{C}]\underset{\textrm{sugra}}{\sim} e^{-S[\mathcal{S}]}
\end{equation}
where $S[\mathcal{C}]$ is the classical Euclidean action of the string world-sheet (that, in fact, does not describe the Wilson loop \eqref{wilsonloop} but instead its supersymmetric generalization). In practice, one usually takes the simplest action which describes the dynamics of an open string, namely the Nambu-Goto action $S_{NG}[\mathcal{C}]=\frac{1}{2\pi\alpha'}\int d^2\xi\sqrt{det(\gamma_{ab})}$ where $\alpha'=\ell_s^2$ is related to the typical length scale $\ell_s$ of the string and $\gamma_{ab}(\xi)$ ($a,b=1,2$) is the induced metric tensor on the two-dimensional world-sheet. Furthermore, it turns out necessary to regularize the potential (by cutting off the infinite range of the holographic coordinate in the \emph{ultraviolet}) as infinities arise: these correspond to the two infinitely stretched strings associated with the infinitely heavy $Q$ and $\bar{Q}$ respectively\footnote{In a flat spacetime, the mass-squared $M^2$ of a string stretched between $x_1$ and $x_2$ reads in broad outline and neglecting the quantum fluctuations as $M^2=\left(\frac{x_2-x_1}{2\pi\alpha'}\right)^2$.}. Hence, the final recipe for computing Wilson loops and regularized potentials is:
\begin{equation}\label{recipe}
V_{Q\bar{Q}}(r)=\lim_{\underset{M\rightarrow\infty}{T\rightarrow\infty}}\frac{1}{T}(S_{NG}-\ell M)
\end{equation} 
from which is derived \eqref{result} and where the second term on the \emph{r.h.s.} consists precisely in subtracting the contribution of the two infinitely stretched $Q$ and $\bar{Q}$ strings ($\ell$ is the total perimeter of the loop $\mathcal{C}$ on the boundary).    

\section{The static potential at finite temperature in supergravity}
Following Hawking and Page's work on the thermodynamics of black holes in anti-de Sitter spacetimes, a gauge/string duality involving a gauge theory at finite temperature was proposed\cite{witten2}: In this framework, the bulk accommodates a Schwarzschild black hole (BH) whose the metric in Euclidean reads as
\begin{equation}\label{line}
ds^2_{BH}=\alpha'\Big\{\frac{u^2}{\tilde{R}^2}\left(f(u)dt^2+d\overrightarrow{x}^2\right)+\frac{\tilde{R}^2}{u^2}\frac{du^2}{f(u)}+\tilde{R}^2d\Omega_5^2\Big\}
\end{equation} 
where $f(u)=1-\frac{u_T^4}{u^4}$ and $\tilde{R}^4\equiv\frac{R^4}{\alpha'^2}=4\pi g_s N=2\lambda$ \footnote{$g_s$ is the closed string coupling constant. Using a standard convention in string theory, the world-sheet coordinates will be written with capital letters.} is the dimensionless $AdS$ radius. $u$ is the holographic coordinate. There is a curvature singularity at $u=0$ hidden behind the event horizon at $u=u_T$ whose the location is given by the Bekenstein-Hawking temperature $T$ of the black hole $u_T=\pi\tilde{R}^2T$. In particular, at zero temperature, \emph{i.e.} when $u_T=0$, one recovers the line element of $AdS_5\times S^5$. To the high (low) energy regime of the dual boundary theory corresponds $u\rightarrow\infty$ ($u\rightarrow0$). Then, considering a \emph{spacetime} Wilson loop (that is, a loop along one space-like dimension and one time-like dimension) and following Maldacena's recipe, the subtracted static potential is\cite{rey}:
\begin{equation}\label{wilson-conformal-potential temperature}
V_{Q\bar{Q}}=\frac{U_0}{\pi}\int_1^{U_{max}/U_0}dv\left(\sqrt{\frac{v^4-1+\epsilon}{v^4-1}}-1\right)+\frac{U_T-U_0}{\pi}
\end{equation}
where $v\equiv U/U_0$ and $\epsilon\equiv f(U_0)$. $U(x)$ is the string coordinate along the fifth holographic dimension which is only a function of the spatial boundary coordinate $x$. Because the string world-sheet is symmetric under the mirror transformation $x\leftrightarrow -x$, $U(x)$ presents a minimum which then occurs at $x=0$. By definition, $U_0\equiv U(0)$ (and $U'(0)=0$). $U_T$ is the value of the string coordinate at the event horizon while $U_{max}$ is the \emph{ultraviolet} cutoff ($U\leq U_{max}$). Let us focus on the limiting case $U_0\gg U_T$ ($\epsilon\simeq1$) where the string world-sheet is close to the boundary such that it does not feel the presence of the horizon. In fact, this
configuration corresponds to the low temperature limit $rT\ll1$. Obviously, for small
temperatures, the potential behaves approximately as in the zero
temperature case $V\sim-\frac{1}{r}$, Eq.\eqref{result}. Moreover, the leading non-zero
temperature correction exhibits scaling consistent with the
conformal invariance of the boundary theory\cite{rey}: $V\propto-\frac{1}{r}\Big(1+a(rT)^4\Big)$ with $a$ a positive numerical constant which does not depend on
$\tilde{R}$. Without length scale, it is indeed meaningless
to speak, at low temperature, of a large or small compactification radius of the Euclidean
temporal dimension (whose the period gives the temperature). The high temperature limit $rT\gg1$ when $U_0\simeq U_T$ is more
subtle. As shown in\cite{rey}, there is a critical value of the inter-quark distance
above which the potential starts to be positive. At this point,
the bound-state potential \eqref{wilson-conformal-potential temperature} is not valid because the lowest energy configuration consists instead
of two straight strings ending at the horizon. In other words, the
quarks become free as screened by the effects of the temperature.
Hence, the potential exhibits behavior expected for the
deconfinement phase at high temperature when the meson decays into
a configuration of quarks without interaction.

As a matter of fact, linear potentials can also appear in supergravity. For this, we will consider \emph{spatial} Wilson loops (\emph{i.e.} along two spacelike dimensions) at \emph{fixed value} of the temperature\cite{witten2}\cite{greensite}. Then, the heavy quark potential and the interquark distance as functions of $U_0$ and $U_T$ are\cite{greensite}:
\begin{equation}
r=\frac{2\tilde{R}^2}{U_0}\int_1^{\infty}\frac{dv}{\sqrt{(v^4-1+\epsilon)(v^4-1)}}\;,\label{exp1}
\end{equation}
\begin{equation}
V_{Q\bar{Q}}=\frac{U_0^2}{2\pi\tilde{R}^2}r+\frac{U_0}{\pi}\int_1^{\infty} dv\left(\sqrt{\frac{v^4-1}{v^4-1+\epsilon}}-1\right)+\frac{U_T-U_0}{\pi}\;.\label{exp2}
\end{equation}
In the limit $U_0\simeq U_T$ ($\epsilon\ll1$) where the string world-sheet
reaches the horizon, the interquark distance diverge, which thus corresponds to the large distance limit. On the other hand, when $rT\gg1$,
the circle $S^1(1/T)$ around the compactified Euclidean time direction is small and, as a
result, the number of dimensions of the $4d$ gauge theory on the boundary reduces to three.
By choosing appropriate boundary conditions along this circle (namely, by taking antiperiodic
fermions around $S^1(1/T)$ in contrast to the periodic bosons), the supersymmetry
can also be broken\cite{witten2}. Moreover, as both fermions and scalars get masses related to
the temperature (due to renormalization for the latter), they decouple at high enough
temperature and the theory reduces to a pure non-conformal gauge theory. We are thus
considering, at large distances, $3d$ non-supersymmetric Yang-Mills theory
at zero temperature. On the contrary, at small distances
$rT\ll1$, the compactification radius of the circle is sizeable. We deal therefore
with $4d$ supersymmetric Yang-Mills theory at zero temperature and, not
surprisingly, we recover Maldacena's result \eqref{result}. An expansion in power series of $\epsilon$ gives the leading and subleading terms in the static potential at large quark separation:
\begin{equation}\label{result2}
V_{Q\bar{Q}}=\frac{U_T^2}{2\pi\tilde{R}^2}r\left(1-\frac{1}{2}e^{-\frac{2U_T}{\tilde{R}^2}r}\right)\;.
\end{equation}
As expected, the string tension is proportional to (the square of) the
temperature since it is our only dimensionful parameter at hand:
\begin{equation}\label{stringtension}
\sigma_T=\frac{U_T^2}{2\pi\tilde{R}^2}=\frac{1}{2}\pi\tilde{R}^2T^2=\sqrt{\pi^3 g_s N}T^2\;.
\end{equation}
The subleading term in the potential \eqref{result2} is not the L\"uscher term in -$1/r$ as predicted
by effective string models and Lattice QCD. The result \eqref{result2} is not so surprising in fact since the
limits at work in the supergravity approach are the large $N$ and the large 't Hooft coupling
constant limits and it is known that there is no L\"uscher term in the strong coupling regime
on the lattice. In this respect, Ref.\cite{greensite} argued that the L\"uscher term could arise from quantum fluctuations of the classical world-sheet approximation. Let us also mention that the linear behavior of the potential is not spoilt by the leading stringy corrections $O(\alpha'^3)$ of the metric \eqref{line}. The expressions \eqref{exp1} and \eqref{exp2} of $r$ and $V_{Q\bar{Q}}$ are modified only by terms in $1/v$ which do not rule out their singular behaviors in the limit $U_0\simeq U_T$.

The Schwarzschild black hole$-AdS_5$ geometry described above was required in order to deal with a $3d$ gauge theory (after compactification of the Euclidean time direction). If we are interested in studying higher-dimensional gauge
theories, it is then necessary to consider the general case of a stack of $N$ coincident Dp-branes in the decoupling limit. We are therefore led
to the (Euclidean) metric\cite{maldacena2}:
\begin{align}
ds^2=&\alpha'\Big\{\frac{u^{\frac{(7-p)}{2}}}{g_{YM}^{(p+1)}\sqrt{d_pN}}(dt^2+d\overrightarrow{x}^2)+\frac{g_{YM}^{(p+1)}\sqrt{d_pN}}{u^{\frac{7-p}{2}}}du^2\nonumber\\
&+g_{YM}^{(p+1)}\sqrt{d_pN}u^{\frac{(p-3)}{2}}d\Omega_{8-p}^2\Big\}\label{wilson-metric general}
\end{align}
with $d_p\equiv 2^{7-2p}\pi^{\frac{9-3p}{2}}\Gamma(\frac{7-p}{2})$ and where the coupling constant $g^{(p+1)}$ of the $(p+1)$-dimensional
$SU(N)$ supersymmetric Yang-Mills theory is related to $g_s$ as ${g^{(p+1)}}^2=(2\pi)^{p-2}g_s\,\alpha'^{\frac{(p-3)}{2}}$. The case of interest here consists of $p=4$, for which ${g^{(5)}}^2=4\pi^2g_s\sqrt{\alpha'}$, and non-zero temperature. The solution of the equations
of motion for the stack of $N$ coincident Dp-branes in the decoupling limit is then:
\begin{equation}
ds^2_{BH}=\alpha'\Big[\frac{u^{3/2}}{R_4^{3/2}}\Big(g(u)dt^2+d\overrightarrow{x}^2\Big)+\frac{R_4^{3/2}}{u^{3/2}}\frac{du^2}{g(u)}+R_4^{3/2}\sqrt{u}\,d\Omega_4^2\Big]
\end{equation}
where $g(u)=1-\frac{u_T^3}{u^3}$ and $u_T=\frac{16}{9}\pi^2R_4^3T^2=\frac{4}{9}\pi{g^{(5)}}^2N\,T^2$ \footnote{We have defined $R_4^{3/2}\equiv
g^{(5)}\sqrt{d_4N}=g^{(5)}\sqrt{\frac{N}{4\pi}}$ such that $R_4$ has the dimension of a \emph{length}$^{1/3}$ ($g^{(5)}$ has
the dimension of a $\emph{length}^{1/2}$).}. In this case, the interquark distance and the static potential read as follows ($\epsilon\equiv
g(U_0)$):
\begin{equation}
r=\frac{2R_4^{3/2}}{U_0^{1/2}}\int_1^{\infty}\frac{dv}{\sqrt{(v^3-1+\epsilon)(v^3-1)}}
\end{equation}
and
\begin{equation}
V_{Q\bar{Q}}=\frac{U_0^{3/2}}{2\pi
R_4^{3/2}}r+\frac{U_0}{\pi}\int_1^{\infty}dv\Big(\sqrt{\frac{v^3-1}{v^3-1+\epsilon}}-1\Big)+\frac{U_T-U_0}{\pi}\;.
\end{equation}
Here also, one can show that the potential presents an area law behavior (when $U_0\simeq U_T$) with a string tension:
\begin{equation}
\sigma_t=\frac{U_T^{3/2}}{2\pi R_4^{3/2}}=\frac{8}{27}\pi
{g^{(4)}}^2N\, T^2
\end{equation}
expressed in terms of the $4d$ gauge theory coupling ${g^{(4)}}^2={g^{(5)}}^2T$. To summarize, we observe an area law for \emph{spatial} Wilson
loops in $4d$ and $5d$ supersymmetric Yang-Mills theories at finite temperature. This is then interpreted as the area law of \emph{spacetime} Wilson loops (after having identified one of the spatial coordinates of the higher-dimensional theory to the non-compactified Euclidean time) in $3d$ and $4d$ non-supersymmetric Yang-Mills theories at zero temperature.

\section{The supergravity description of baryons}

In a $SU(N)$ Yang-Mills theory, a color-singlet baryon must be
made of $N$ quarks. As described in the supergravity dual, such a
baryon consists of $N$ quarks living on the boundary. On each of these quarks ends a string
whose the other endpoint is attached to a D5-brane wrapped around $S^5$: the so-called \emph{baryon vertex} located at the
holographic coordinate $u_0$\cite{wittenbaryon}.
The typical radius of the baryon is denoted $r$. Moreover,
the configuration of the $N$ quarks on the boundary is symmetric
with respect to the boundary dimensions such that the resulting
force acting on the \emph{baryon vertex} is zero along these directions.
In the following, we will consider only the induced metric
contribution in the Dirac-Born-Infeld action of the
D5-brane\cite{baryon}\footnote{In general, Dp-branes carry electromagnetic
fields on their $(p+1)$-dimensional world-volume whose the
dynamics is governed by the so-called Dirac-Born-Infeld action:
$S_{Dp}=T_p\int d^{p+1}x\sqrt{-det(\eta_{MN}+2\pi\alpha' F_{MN})}$
with $T_p=\frac{2\pi}{(2\pi l_s)^{p+1}g_s}$ the brane tension and
$M,N=0,1,\ldots,p$ the spacetime indices on the (flat) world-volume
of the Dp-brane. Especially, $T_5^{-1}=(2\pi)^5{\alpha'}^3g_s$.}:
\begin{equation}\label{wilson-baryons-brane action}
S_{D5}=T_5\int d^6x\sqrt{det\,g_{D5}}=\frac{T\,N\,U_0}{8\pi}\;.
\end{equation}
The total action of the baryonic system is thus:
\begin{align}
S_{total}=&S_{D5}+\sum_{i=1}^N
S^{(i)}_{string}\nonumber\\
=&\frac{T\,N\,U_0}{8\pi}+\frac{T\,N}{2\pi}\int_0^rdx\sqrt{{U'}^2+\frac{U^4}{\tilde{R}^4}}
\end{align}
where the integral over the boundary spatial coordinate $x$ runs from 0 to the typical radius $r$ of the baryon. The stability
(or no-force) condition for the \emph{baryon vertex} along the holographic coordinate stems from variational principle and reads:
\begin{equation}\label{wilson-stability condition}
\delta S_{total}|_{\underset{\textrm{term at }U_0}{\textrm{surface
}}}=0\;\;\Rightarrow\;\;\frac{U'_0}{\sqrt{{U'_0}^2+\frac{U_0^4}{\tilde{R}^4}}}=\frac{1}{4}\;.
\end{equation}
On the other hand, Maldacena's recipe gives the energy of the baryon:
\begin{equation}
V_B(r)=-N\alpha_B\frac{\sqrt{2\lambda}}{r}
\end{equation}
with a coefficient given by
\begin{equation}
\alpha_B=\frac{1}{2\pi}\int_1^{\infty}\frac{du}{u^2\sqrt{\beta^2u^4-1}}\Big\{\frac{3}{4}-\int_1^{\infty}dv\Big[\frac{\beta
v^2}{\sqrt{\beta^2v^4-1}}-1\Big]\Big\}
\end{equation}
of numerical value $\alpha_B\simeq0.036$ ($\beta=\sqrt{\frac{16}{15}}$) and whose the behavior in $1/r$ is dictated by the conformal invariance of the field theory on the
boundary.

Remarkably, another string configuration has been identified\cite{baryon} which allows, on the supergravity
side, to account for baryons made of a smaller number of quark constituents
$k < N$. In such a configuration, $k$ strings attached to the \emph{baryon vertex} end on the $k$ quarks at the boundary while the $N-k$ remaining strings stretch out up to the brane at $u=0$. The total action governing the dynamics of the baryon is then:
\begin{align}\label{wilson-CFT baryon action}
S_{total}=&S_{D5}+\sum_{i=1}^{k}S^{(i)}_{string}+\sum_{j=1}^{N-k}S^{(j)}_{string}\nonumber\\
=&\frac{T\,N\,U_0}{8\pi}+\frac{k\,T}{2\pi}\int_0^{r}dx
\sqrt{{U'}^2+\frac{U^4}{\tilde{R}^4}}+\frac{T(N-k)U_0}{2\pi}
\end{align}
with the following stability
condition for the \emph{baryon vertex} along the holographic coordinate:
\begin{equation}
\delta S_{total}|_{\underset{\textrm{term at }U_0}{\textrm{surface
}}}=0\;\;\Rightarrow\;\;\frac{U_0'}{\sqrt{{U_0'}^2+\frac{U_0^4}{\tilde{R}^4}}}=\frac{5N-4k}{4k}\equiv
A\;.
\end{equation}
If $k=N$, then $A=\frac{1}{4}$ and we recover
\eqref{wilson-stability condition}. The upper
bound for $A$ (which corresponds to the lower bound for $k$) is
obtained for radial straight $k$-type strings ending on the \emph{baryon
vertex} such that $U_0'\to\infty$. Then, $A=1$ or $k=\frac{5N}{8}$.
Finally, the condition for having a stable string-brane
system into the bulk demands $\frac{5N}{8}\leq k\leq N$. As for the potential, if
$k=\frac{5N}{8}$ then $V_B(U_0)=0$ independently of the
location $U_0$ of the \emph{baryon vertex} along the holographic coordinate (actually, the size $r(U_0)$ of the baryon vanishes).
If $\frac{5N}{8}<k\leq N$ then the energy
$V_B(r)=-\alpha\,U_0(r)$ can be written as the product of a
negative constant $-\alpha$ with $U_0$ expressed in terms of the baryon radius
$r$\cite{baryon}.

To conclude this section, let us consider a \emph{spatial} string/brane configuration
in the Schwarzschild black hole$-AdS_5$ background \eqref{line}. Following the same techniques as described above, the total action as well as the stability condition of the system can be derived. We are interested in the large distance regime where the typical radius of the baryons is large and where the D5-brane reaches the
horizon ($U_0\simeq U_T$). A linear potential then arises with a string tension equals to $N$
times the mesonic string tension \eqref{stringtension}:
\begin{equation}
V_B(r)=N\Big(\frac{1}{2}\pi\tilde{R}^2T^2\Big)r\;.
\end{equation}

\section{The heavy quark potential in holographic models of QCD}

\subsection{The heavy quark potential from general geometry in $AdS$/QCD}

We consider a general form of the static
metric which respects Poincar\'e symmetry on the boundary\cite{white}:
\begin{equation}\label{wilson-background metric}
ds^2=\alpha'\tilde{R}^2\Big(f(z)\delta_{\mu\nu}dx^{\mu}
dx^{\nu}+\frac{dz^2}{z^2}\Big)
\end{equation}
with $\delta_{\mu\nu}=\textrm{diag}(+1,+1,+1,+1)$ the $4d$ Euclidean flat metric. The warp factor $f(z)>0$ is assumed to be
positive. In particular, $f(z)=\frac{1}{z^2}$ corresponds to the
Euclidean $AdS_5$ line element in Poincar\'e coordinates\footnote{The change of holographic coordinate $z=\frac{R^2}{\alpha' u}=\frac{\tilde{R}^2 }{u}$ gives the line element \eqref{line} with $u_T=0$.}. The Nambu-Goto action (in the
static gauge where $X^0(\tau,\sigma)=\tau$ and $\sigma=x$) reads
\begin{equation}
S_{NG}[\mathcal{C}]=\frac{\tilde{R}^{2}}{2\pi}\int_{-T/2}^{T/2}dt\int_{-r/2}^{r/2}dx\,f(Z)\sqrt{1+\frac{{Z'}^2}{f(Z)Z^2}}
\end{equation}
where $Z(x)$ is the holographic coordinate of the string. The
Lagrangian density does not depend explicitly on $x$ which gives the first integral:
\begin{equation}
\frac{f(Z)}{\sqrt{1+\frac{{Z'}^2}{f(Z)Z^2}}}=f_0
\end{equation}
where $Z_0$ is the value of $Z(x)$ at $x=0$, that is, the maximal
extend of the string world-sheet along the holographic dimension
where $Z'(0)=0$ by symmetry and $f_0\equiv f(Z_0)$. 

The interquark distance takes the general form:
\begin{equation}
r(Z_0)=2\int_0^{Z_0}dZ\frac{1}{\sqrt{\tilde{f}}}\Big(\frac{Z_0^4\tilde{f}^2}{Z^4\tilde{f}_0^2}-1\Big)^{-\frac{1}{2}}\;.\label{wilson-white-eq1}
\end{equation}
Following\cite{white}, one defines $\tilde{f}(Z)=Z^2\,f(Z)$
such that $\tilde{f}(0)=1$. As for the interaction potential, we have:
\begin{equation}
V(Z_0)=\frac{\tilde{R}^2}{\pi}\Big\{-\frac{1}{Z_0}+\int_0^{Z_0}\frac{dZ}{Z^2}\Big[\sqrt{\tilde{f}}\Big(1-\frac{Z^4\tilde{f}_0^2}{Z_0^4\tilde{f}^2}\Big)^{-\frac{1}{2}}-1\Big]\Big\}\label{wilson-white-eq2}
\end{equation}
which is obtained as usual, once the
infinite contribution $\frac{\tilde{R}^2}{\pi}\frac{1}{Z_{min}}$
($Z_{min}\to0$) stemming from the infinitely stretched strings associated with
the infinitely massive quarks is subtracted.

At short distances, \emph{i.e.} when the string
world-sheet is close enough to the boundary spacetime, the bulk
geometry felt by the string is nearly $AdS_5$. Not surprisingly,
the limit $Z_0\to0$ (and then $Z\to0$ and $\tilde{f}(Z)\to1$ in
\eqref{wilson-white-eq1} and \eqref{wilson-white-eq2}) gives the
famous AdS/CFT result \eqref{result}:
\begin{align}
r(Z_0)&\underset{Z_0\to0}{\simeq}\frac{Z_0}{\rho}\;,\\
V(Z_0)&\underset{Z_0\to0}{\simeq}-\frac{\tilde{R}^2}{2\pi\rho}\frac{1}{Z_0}
\end{align}
such that ($\rho=\frac{\Gamma(1/4)^2}{(2\pi)^{3/2}}$)
\begin{equation}
V(r)=-\frac{\tilde{R}^2}{2\pi\rho^2}\frac{1}{r}\;.
\end{equation}

On the other hand, in the case of mesonic bound-states, the
confinement criterion can be stated as follows: there exists a
finite value $Z_0^{\ast}$ of the maximal extent of the world-sheet
along the holographic coordinate such that the interquark distance
$r(Z_0^{\ast})$ diverges. This peculiar value
$Z_0^{\ast}$ is related to the QCD mass gap. In particular, it
enters the expression of the string tension in the linear potential. Moreover, this divergence is logarithmic. By
expanding around $Z_0^{\ast}$, we have indeed:
\begin{equation}
r(Z_0^{\ast})\underset{Z\to
Z_0^{\ast}}{\sim}-\ln(1-\frac{Z}{Z_0^{\ast}})
\end{equation}
provided the following relation holds (where $\tilde{f}(Z_0^{\ast})\equiv\tilde{f}_0^{\ast}$):
\begin{equation}
Z_0^{\ast}\frac{d\tilde{f}}{dZ}\Big|_{Z_0^{\ast}}=2\tilde{f}_0^{\ast}
\end{equation}
which establishes the confinement criterion\footnote{The issue of finding general criteria for the confinement has also
been considered in\cite{kiritsis}.}. Finally, the static potential reads
\begin{equation}
V(r)=\sigma_t(Z_0^{\ast})\,r
\end{equation}
with the string tension:
\begin{equation}
\sigma_t(Z_0^{\ast})=\frac{\tilde{R}^2}{2\pi}\frac{\tilde{f}_0^{\ast}}{Z_0^{\ast\,2}}\;.
\end{equation}

\subsection{Andreev and Zakharov's model}

As an explicit holographic models of QCD, we turn our attention to Andreev and Zakharov's model\cite{AZ}. As always in $AdS$/QCD, it implies to introduce a dimensionful parameter related in some way to the QCD mass gap. This can be, for instance, the cutoff $z_m$ of the $AdS_5$ slice as in the Hard Wall model or the dilaton parameter in the Soft Wall Model\cite{models}. In Ref.\cite{AZ}, the isometry
group of the holographic spacetime $AdS_5$ (\emph{i.e.} the
conformal invariance of the boundary field theory) is broken by means of a
warp factor $h(z)$ in the Euclidean metric:
\begin{equation}\label{wilson-AZ line element}
ds^2=\frac{R^2}{z^2}h(z)\,\delta_{MN}dx^Mdx^N\;.
\end{equation}
The bulk coordinates are $x^M=(x^{\mu},z)$ with $x^{\mu}$
$(\mu=0,\ldots,3)$ the boundary coordinates and $z>0$ is the
holographic coordinate. $\delta_{MN}=\textrm{diag}(+1,+1,+1,+1)$ is
the Euclidean flat metric. In this model, the warp factor
$h(z)\equiv e^{\frac{1}{2}c\,z^2}$ introduces the conformal
symmetry breaking parameter $c$ and we recover the
$AdS_5$ metric \eqref{wilson-background metric} with $f(z)=\frac{1}{z^2}$ near the
\emph{ultraviolet} brane $z\to0$ where $h(0)=1$. As usual, we start from the Nambu-Goto action in the static gauge: 
\begin{equation}
S_{NG}[\mathcal{C}]=\frac{\tilde{R}^2}{2\pi}T\int_{-\frac{r}{2}}^{\frac{r}{2}}dx\frac{h}{Z^2}\sqrt{1+{Z'}^2}
\end{equation}
where the holographic coordinate of the string $Z(x)$ is only a
function of $x$ and $\mathcal{C}$ is the rectangular loop of sides $r$ and $T$. The two parametric expressions for the interquark distance
$r(Z_0,c)$ and the interaction potential $V(Z_0,c)$ take the
following forms ($Z(0)=Z_0$ and $Z'(0)=0$ by symmetry. $\lambda=c Z_0^2$ and $v=Z/Z_0$):
\begin{align}
r=&2\sqrt{\frac{\lambda}{c}}\int_0^1dv\,v^2\,e^{\frac{1}{2}\lambda(1-v^2)}\Big(1-v^4e^{\lambda(1-v^2)}\Big)^{-\frac{1}{2}}\;,\label{wilson-AZ interquark distance}\\
V=&\frac{\tilde{R}^2}{\pi}\sqrt{\frac{c}{\lambda}}\int_0^1
dv\frac{e^{\frac{1}{2}\lambda\,v^2}}{v^2}\Big(1-v^4\,e^{\lambda(1-v^2)}\Big)^{-\frac{1}{2}}\;.\label{wilson-AZ
divergent integral potential}
\end{align}
As expected, the integral
\eqref{wilson-AZ divergent integral potential} does not converge
when $v\to0$ and require a \emph{ultraviolet} cutoff $Z(x)\geq Z_{min}$. Finally, the substracted potential reads:
\begin{equation}\label{wilson-AZ renormalized potential}
V=\frac{\tilde{R}^2}{\pi}\sqrt{\frac{c}{\lambda}}\Big\{-1+\int_0^1\frac{dv}{v^2}\Big[e^{\frac{1}{2}\lambda\,v^2}\Big(1-v^4\,e^{\lambda(1-v^2)}\Big)^{-\frac{1}{2}}-1\Big]\Big\}\;.
\end{equation}
The expression \eqref{wilson-AZ interquark
distance} has a logarithmic singularity when $\lambda=c\,Z_0^2=2$. This
peculiar finite value of $Z_0=\sqrt{\frac{2}{c}}$ corresponds to
the maximal extent reached by the string along the
holographic coordinate. There, the interquark distance diverges, which mimics the confinement mechanism. On the contrary,
if the conformal symmetry parameter $c=0$, then $Z_0$ is allowed
to run all over the holographic dimension and we have no longer confinement. Let us identify this logarithmic
singularity. As we are interested in the region $Z\sim
Z_0$, \emph{i.e.} $v\sim1$, the integral in \eqref{wilson-AZ interquark distance} can be
approximately replaced by its main contribution:
\begin{equation}
r\propto\int_0^1\frac{dv}{\sqrt{2(2-\lambda)(1-v)+(-2\lambda^2+9\lambda-6)(1-v)^2}}
\end{equation}
which clearly has a logarithmic singularity at $\lambda=2$. The static potential \eqref{wilson-AZ renormalized potential}
develops the same singularity when $v\to1$ at $\lambda=2$ and, as such, is linear at long
distances:
\begin{equation}
V(r,c)=\sigma_t r
\end{equation}
with the long-distance string tension:
\begin{equation}\label{wilson-AZ long distance string tension}
\sigma_t=\tilde{R}^2\frac{e}{4\pi}c\;.
\end{equation}

At short distances, the behaviors of $r$ and $V$ correspond to a string configuration with $Z_0\sim0$, namely to
the limit $\lambda\to0$. Then, the heavy quark potential reads as
\begin{equation}
V(r)=-\frac{\kappa_0}{r}+\sigma_0\,r+O(r^3)\label{wilson-AZ potential}
\end{equation}
with
\begin{equation}
\left\{
\begin{array}{lll}
\kappa_0&=&\frac{\tilde{R}^2}{2\pi\rho^2}=\frac{\sqrt{4\pi g_s N}}{2\pi\rho^2}\;\;,\\
\sigma_0&=&\tilde{R}^2\frac{c\,\rho^2}{4}\;\;.\label{wilson-AZ short
distance string tension}
\end{array}
\right.
\end{equation}
It appears, in the supergravity side, two
tensions $\sigma_t$ \eqref{wilson-AZ long distance string tension} and
$\sigma_0$ \eqref{wilson-AZ short distance string tension}
corresponding respectively to the long and to short distance
regimes. Nevertheless, their ratio turns out to be rather closed
to one:
\begin{equation}
\frac{\sigma_t}{\sigma_0}=\frac{e}{\pi\rho^2}=\frac{8\pi^2 e}{\Gamma(1/4)^4}\simeq1.24\;,
\end{equation}
an estimate satisfactory at the usual accuracy level of the holographic models
of QCD. As for the Coulomb-like term in \eqref{wilson-AZ potential}, it does not have to be identified with the
perturbative part of the Cornell potential. This is reminiscent of the $AdS$/CFT result \eqref{result}.

\acknowledgments{It is a pleasure to thank the organizers of the conference for collaborations and discussions. I am grateful to Conselho Nacional de Desenvolvimento Cient\'ifico e Tecnol\'ogico CNPq for financial support (grant 150252/2011-0).} 

\end{multicols}

\vspace{-2mm}
\centerline{\rule{80mm}{0.1pt}}
\vspace{2mm}

\begin{multicols}{2}

\end{multicols}

\clearpage

\end{document}